# Tunable Transmissive Metagratings Using Single Layer Cylindrical Plasma Discharges

Mohammad G. H. Alijani, Alessio Monti, *Senior Member, IEEE*, Stefano Vellucci, *Member IEEE*, Mirko Barbuto, *Senior Member, IEEE*, Alessandro Toscano, *Senior Member, IEEE*, Filiberto Bilotti, *Fellow, IEEE*

***Abstract*— In this paper, we propose a novel single-layer reconfigurable transmissive metagrating based on plasma discharges. Each unit-cell consists of two side-by-side core–shell cylinders, with a tunable plasma core and a high-index dielectric shell. The structure is modeled using a free-electron plasma permittivity with adjustable plasma frequency. Analytical and numerical results show that the main transmission lobe can be switched between −41°, 0° and 41° by tuning the plasma frequencies. Transmission efficiency remains above 80% at broadside and 90% in the steered direction. This tunability enables effective directional control with low reflection and high overall power efficiency.**

## I. INTRODUCTION

Metagratings are periodic structures that control electromagnetic waves by engineering diffraction rather than relying on refraction. When excited by an incident wave, they support multiple diffracted modes, whose directions are determined by the grating equation. By properly designing the unit-cell—through geometry and material parameters—specific modes can be suppressed or enhanced, enabling efficient beam steering and shaping in both transmission and reflection [1]-[3]. A variety of design strategies have been proposed [4]-[5], involving nonlocal responses [6]-[7], aperiodic arrangements [8], and targeting diverse functionalities such as beamforming, beam splitting, and wavefront shaping [9]-[12].

Recent efforts have focused on reconfigurable metagratings, capable of dynamically modifying their response through external stimuli [13]-[14]. Common tuning strategies rely on electronic or material-based mechanisms (e.g., varactors, MEMS, PIN diodes), but these often face limitations in terms of losses, fabrication complexity, discrete tuning, or limited bandwidth. In this context, plasma discharges offer a competitive alternative in certain operating scenarios. They allow for continuous tunability, MHz-level switching speeds, high-frequency operation, low cost, and high-power robustness. Moreover, they can be turned off when not needed and may allow post-fabrication tuning of the operating frequency. Plasma-based elements have already been used in antennas [15]-[18], metamaterials [19], EBGs [20]-[21], topological structures [22][23], and reconfigurable surfaces [24]-[26].

Building upon the results obtained [27]-[28], where we demonstrated the potential of multilayer plasma-based gratings, here we show that a single-layer transmissive metagrating can achieve similar beam-switching performance by coating each plasma cylinder with a high-permittivity dielectric shell. This is enabled by the interplay of multiple resonances in the core–shell structure. Theoretical predictions, supported by full-wave simulations, confirm the scalability and reconfigurability of this approach for advanced beam control.

## II. ANALYTICAL MODELLING OF THE UNIT-CELL

The metagrating geometry proposed in this paper is shown in Fig. 1. Each unit-cell consists of two side-by-side core-shell cylinders, separated by distance $d$. In the theoretical formulation, it is assumed that the cylinders are infinitely long; however, in practical applications, it is sufficient that their length is higher than the operation wavelength. The inner and outer radii of each cylinder are equal to $r_i$ and $r_o$, respectively. The cylinders are formed by a plasma core enclosed within a dielectric container with relative dielectric permittivity $\varepsilon_r$. The repetition period of the metagrating is denoted with $P$. As shown in Fig. 1, the metagrating is illuminated by a plane wave with free space propagation constant $k_0$ and $TM^z$ polarization.

As it is well known, the relative dielectric permittivity $\varepsilon_p$ of plasma discharges is well-described by the Drude model [29], *i.e.,*

$$\varepsilon_p^i(\omega) = 1 - \frac{\left(\omega_p^i\right)^2}{\omega\left(\omega - j\upsilon_p^i\right)}, \quad i = 1,2 \tag{1}$$

where, $\omega$ is the angular frequency of the impinging wave, whereas $\omega_p$ and $\upsilon_p$ represent the plasma parameters, *i.e.* the plasma frequency and the collision frequency. The plasma frequency depends primarily on the electron density and electron mass and can be changed within a wide range of possible values [30], while the collision frequency is a function

Manuscript received March 3, 2025; revised March 20, 2025; accepted May 10, 2025. Date of publication May 20, 2025; date of current version March 15, 2025. This work has been developed in the frame of the activities of the Project PULSE, funded by the European Innovation Council under the EIC Pathfinder Open 2022 program (protocol number 10109931).
(Corresponding author: Alessio Monti.)

M. G. H. Alijani, A. Monti, M. Barbuto, A. Toscano, and F. Bilotti are with the Department of Industrial, Electronic and Mechanical Engineering, Roma Tre University, 00146 Rome, Italy (e-mail: alessio.monti@uniroma3.it). S. Vellucci is with the Department of Engineering, Niccolò Cusano University, Via don Carlo Gnocchi 3, 00166, Rome, Italy. M. Barbuto F. Bilotti and S. Vellucci are also with the Virtual Institute for Artificial Electromagnetic Materials and Metamaterials, Louvain-la-Neuve, Belgium.

of particle density, temperature, particle mass, and collision cross-section [29].

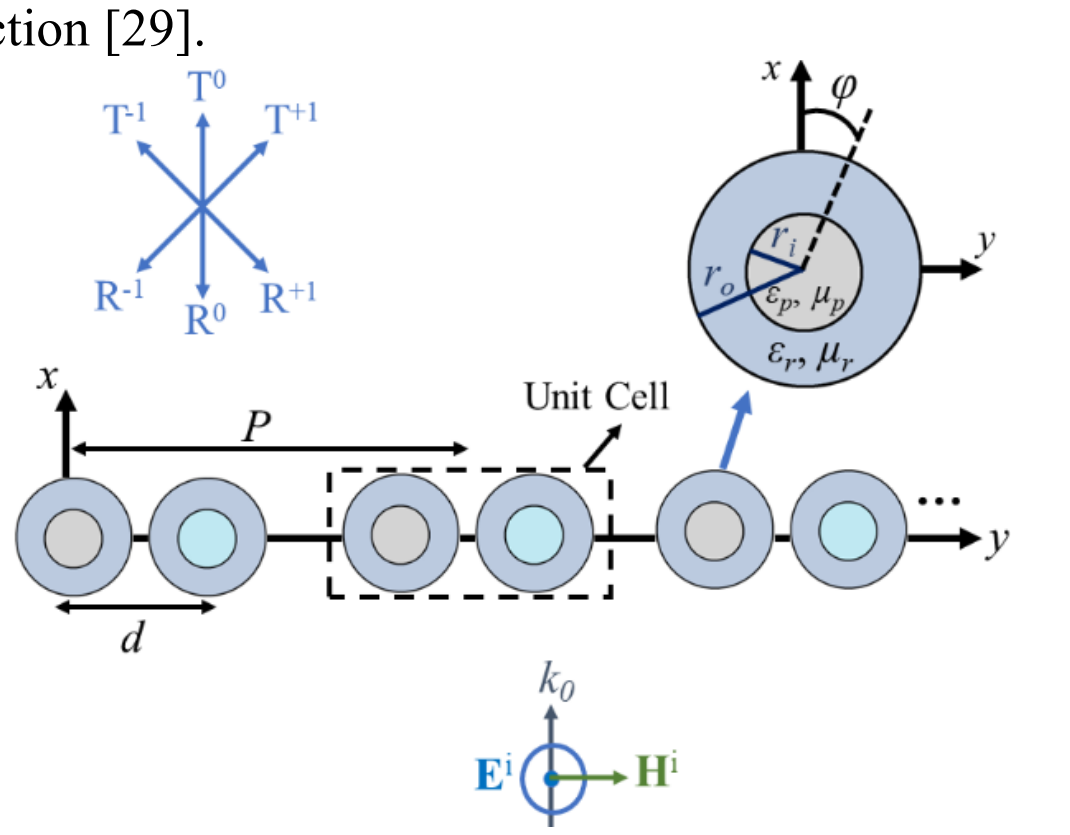

**Fig. 1.** Sketch and geometrical configuration of the proposed reconfigurable metagrating (top view) illuminated by $TM^z$ plane wave.

The scope of this section is to outline an analytical approach to determine the scattering pattern of a metagrating consisting of a finite number $N$ of unit-cells. To this end, we employ cylindrical harmonic expansions, which are well-suited to model the core–shell plasma-dielectric cylinders used as scatterers. Indeed, unlike line-current models adopted for instance in [5],[31], the scatterers considered here exhibit a complex multipolar response that also depends on the excitation voltage and requires a more accurate representation. The proposed approach also enables precise modeling of finite-size metagratings, differently from Floquet-based methods that assume infinite periodicity.

In the scenario depicted in Fig. 1, the electric field of the incident wave $E^{inc}_z$ can be expanded using the cylindrical Hankel function of the first and second kind [32], *i.e.,*

$$E_z^{inc} = e^{-jk_0\rho\cos\varphi} = \frac{1}{2}\sum_{m=-\infty}^{+\infty}\left(-j^m\right)\left[H_m^{(1)}\left(k_0\rho\right)+H_m^{(2)}\left(k_0\rho\right)\right]e^{jm\varphi} \quad (2)$$

Similarly, for an isolated core-shell plasma cylinder, the electric field $E_z^{core}$ inside the core, the electric field $E_z^{shell}$ inside the dielectric shell and the scattered electric field $E_z^{s}$ can be expanded as follows:

$$\begin{aligned} E_z^{core} &= \sum_{m=-\infty}^{+\infty} D_m\left[H_m^{(1)}\left(k_i\rho\right)+H_m^{(2)}\left(k_i\rho\right)\right]e^{jm\varphi} && \rho < r_i \\ E_z^{shell} &= \sum_{m=-\infty}^{+\infty}\left[B_m H_m^{(1)}\left(k_o\rho\right)+C_m H_m^{(2)}\left(k_o\rho\right)\right]e^{jm\varphi} && r_i<\rho<r_o \quad (3) \\ E_z^{s} &= E_z^{inc} + \sum_{m=-\infty}^{+\infty} A_m H_m^{(2)}\left(k_0\rho\right)e^{jm\varphi} && \rho > r_0 \end{aligned}$$

where $k_i=\omega(\mu_p\varepsilon_p)^{1/2}$ and $k_o=\omega(\mu_o\varepsilon_r)^{1/2}$ are the propagation constant of the waves inside the core and the shell, respectively, whereas $A_m$, $B_m$, $C_m$ and $D_m$ are the $m^{th}$ harmonic coefficient of the scattering mode. The magnetic field at each region can be directly calculated using the first curl Maxwell equation. It is worth mentioning that the electric fields in Eq. (3) are normalized to the cylinder length $L$ to facilitate comparison across different geometries or configurations. By matching the tangential components of the electric and magnetic fields at the interfaces of the core-shell cylinder $E_z^{core}(\rho=r_i) = E_z^{shell}(\rho=r_i)$, $E_z^{shell}(\rho=r_o) = E_z^{s}(\rho=r_o)$, $H_\varphi^{core}(\rho=r_i) = H_\varphi^{shell}(\rho=r_i)$, and $H_\varphi^{shell}(\rho=r_o) = H_\varphi^{s}(\rho=r_o)$, a system of four equations with four unknowns is established, which by solving it, the four unknowns in Eq. (3) can be determined [32].

In the unit cell of the metagrating shown in Fig. 1, the field scattered by each cylinder acts as an additional incident wave on the others. Therefore, mutual coupling between cylinders must be taken into account. Since the scattered fields are expanded in terms of Hankel functions of the second kind, as in Eq. (3), Graf's addition theorem can be applied to account for inter-element interactions. Further details on this procedure can be found in [33]. To this end, the harmonic coefficients of the scattered field ($A_m$) for $\rho>r_o$ should be replaced with $a_m$ defined as:

$$a_m = \sum_{q=1}^{\infty} H_{m-q}^{(2)}\left(qk_0d\right)\left[e^{jk_0qd}+\left(-1\right)^{m-q}e^{-jk_0qd}\right] \quad (4)$$

where $d$ is the two cylinders spacing.

Finally, the total scattered field of the metagrating $E^s_{z,total}$ with a finite number of unit-cells $N$ can be calculated using the array theory:

$$E^s_{z,\text{total}} = \sum_{n=1}^{N}\sum_{m=-\infty}^{+\infty} a_{mn}H_m^{(2)}\left[k_0\rho_n\right]e^{jm\varphi_n} \quad (5)$$

where $\rho_n=(x^2+(y-nP)^2)^{1/2}$, $\sin\varphi_n = (y-nP)/\rho_n$, $x = \rho\cos\varphi$ and $y=\rho\sin\varphi$. Our results – not shown here for the sake of brevity - confirm that, by considering the $k_0r_i$+1 number of harmonics, agreement with full-wave simulations is acceptable, as also shown in [33]. It is worth noticing that, although this work primarily focuses on $TM^z$ polarization—since, as will become evident later, it enables effective engineering of the unit-cell scattering—preliminary results for $TE^z$ polarization in a metasurface configuration are available in [26].

It is worth noting that no explicit plane-wave conversion of the scattered field is required in this formulation. Instead, the design is carried out directly in the cylindrical harmonic domain by enforcing a maximum in the desired grating direction and nulls in the others, as will be shown in the next section.

## III. Analys and Design of the Metagrating

The periodic structure shown in Fig. 1 excites multiple diffraction modes whose propagation direction can be determined through the grating equation [28], *i.e.,* $\sin(\varphi_t) = \pm t\lambda_0/P$, $t$=0,±1,…, where $\varphi_t$ is the angle of the $t$-th diffracted wave relative to the broadside direction. Our goal is to employ the analytical model developed above to achieve a preliminary design of a transmissive metagrating capable of switching the transmitted field among three directions: $\varphi_t$=+41°, $\varphi_t$=−41°, and $\varphi_t$ =0°. To this end, we first set the metagrating period $P$ to support these angles in the transmission mode, *i.e.,* $P$=1.5$\lambda_0$, and then optimize the available degrees of freedom—namely, the shell permittivity and the plasma frequencies of the two cylinders. The design is structured in two steps: first, a parametric study is carried out to get some physical insights and to investigate how different combinations of shell permittivity and plasma frequencies affect the main lobe direction and efficiency; then, based on these insights, a multi-objective genetic algorithm is applied directly to the analytical model to identify optimal parameters.

Fig. 2 illustrates the effect of the plasma frequencies of the two cylinders on the direction $\varphi_{max}$ of the maximum radiation on the horizontal plane. Geometrical parameters are $d$=0.43$\lambda_0$, $P$=1.5$\lambda_0$, $r_1$=0.09$\lambda_0$, $r_2$=0.10$\lambda_0$. Please note that the plasma frequencies values are normalized to the frequency of the incident wave $f_0$ and that the results are presented for different values of the shell permittivity (*i.e.,* $\varepsilon_r$=3 and $\varepsilon_r$=19). In addition, it is worth noting that although this grating supports six diffraction orders, the dominance of a single direction in Fig. 2 arises from the engineered unit-cell scattering, which enables the selective suppression or enhancement of specific modes. These plots confirm that, for both the shell permittivities, the main lobe direction can be controlled by changing the plasma frequencies of two plasma rods. In other words, a proper plasma frequency pair enables the selective suppression of undesired diffraction modes. Moreover, when the shell permittivity is higher, the tuning process becomes more flexible, as a wider dynamic range of plasma frequencies can be exploited.

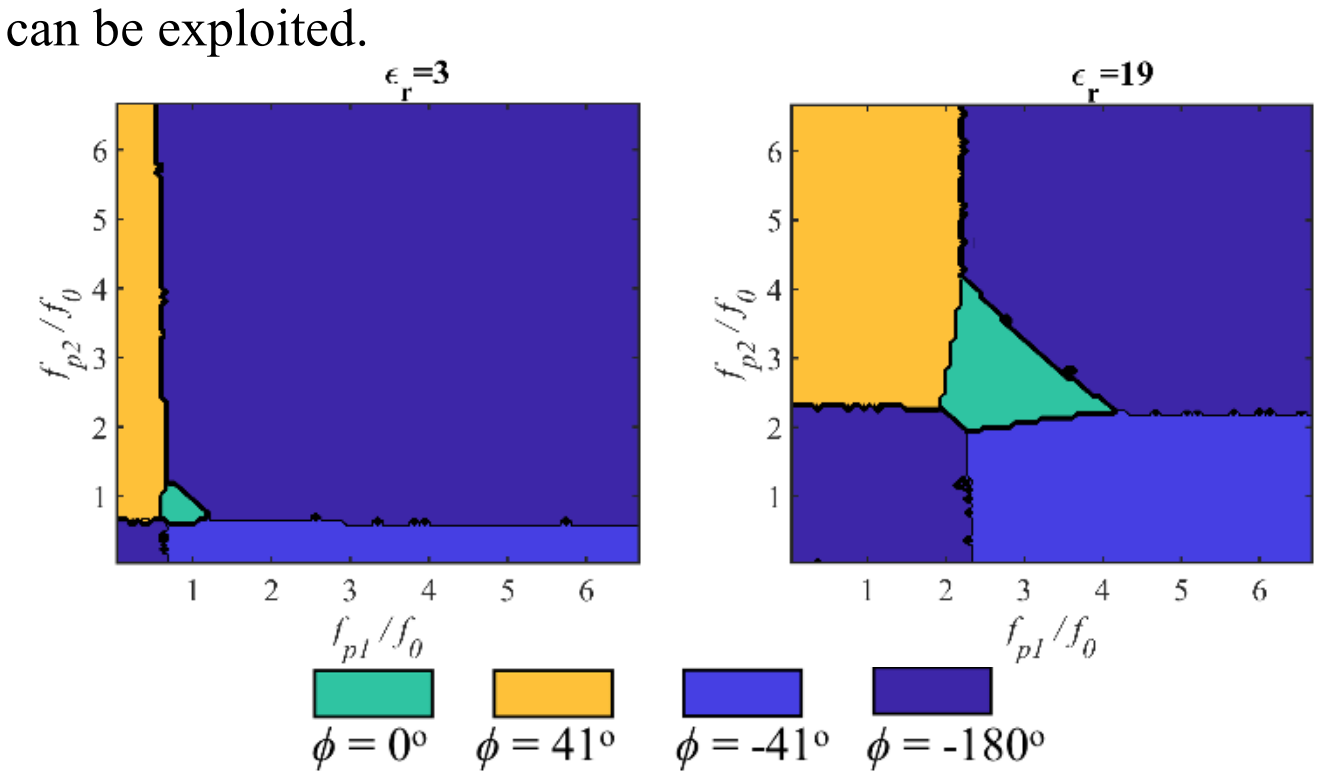

**Fig. 2.** Main lobe direction of the metagrating for different combinations of plasma frequencies of the two cylinders. The left (right) panel refers to a dielectric shell with $\varepsilon_r$=3 ($\varepsilon_r$=19).

As a second step, we focus on the transmission efficiency for different main lobe directions. According to the results in Fig. 2, we consider three different scenarios: broadside transmission (*i.e.,* $\varphi_t$ = 0°), positive steered beam (*i.e.,* $\varphi_t$ = 41°), and negative steered beam (*i.e.,* $\varphi_t$ = -41°). The corresponding transmission coefficients, computed through full-wave simulations in unit-cell environment, are denoted with $T^0$ and $T^{\pm1}$, respectively, and are reported in Fig. 3 vs. the shell permittivity ($f_{p1}/f_0$=4.97, $f_{p2}/f_0$=0.17). In addition, the figure shows the behavior of the undesired scattering level (USL), defined as the relative amplitude of the second-highest peak of the scattered electric field—typically corresponding to a non-targeted Floquet transmission order—with respect to the main lobe (i.e., the desired diffraction direction). While analogous to grating lobe levels in antenna arrays, USL refers to a scattering problem and specifically quantifies diffraction into undesired orders. The USL has been calculated through full-wave simulations in CST using unit-cell boundary conditions and then applying the array factor to model a finite grating of 10 elements. As can be appreciated from Fig. 3, the choice of the optimal shell permittivity is a trade-off between different requirements. For broadside beam, low permittivity ensures high transmission and low USL. Indeed, in such a scenario, the metagrating unit-cell should interact as minimally as possible with the impinging wave to avoid exciting diffraction modes. On the contrary, low-permittivity values return a magnitude of $T^{\pm1}$ below 0.5, showing unsatisfactory cancellation of reflection modes in the steered scenarios. A reasonable trade-off is achieved for a value of permittivity around 20: in that case, the transmission coefficient $T^{\pm1}$ is maximized, while the value of $T^0$ is above 0.8.

The improved transmission in the steered case stems from the interplay between magnetic dipole resonances in the high-permittivity shell and electric dipole resonances in the plasma core. A low-index dielectric does not support magnetic resonance and, as such, does not allow enough degrees of freedom for scattering pattern engineering [2].

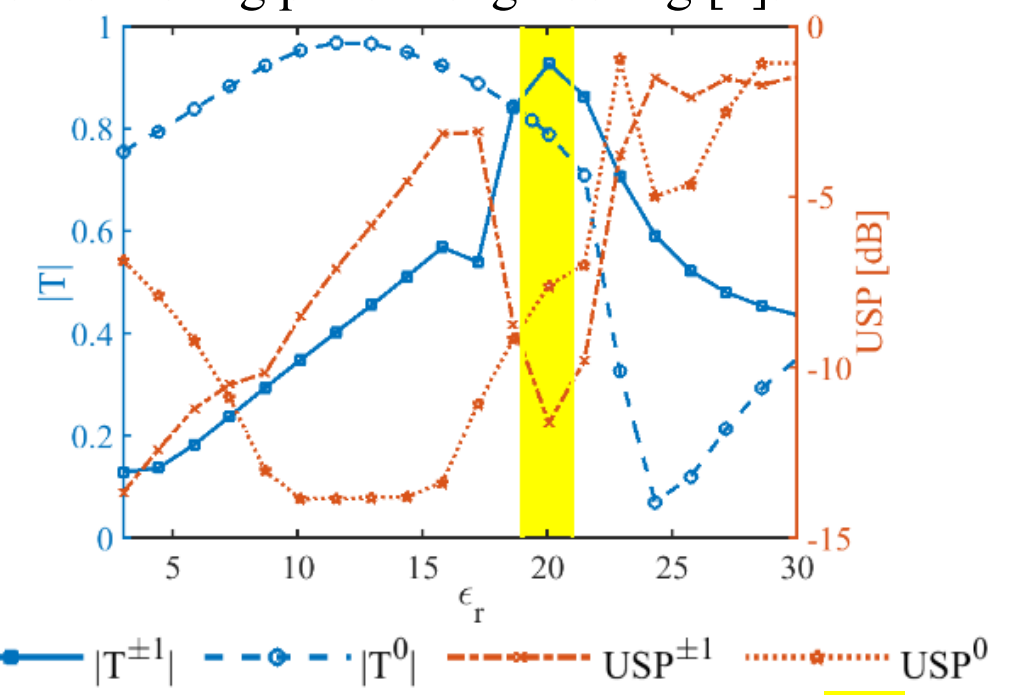

**Fig. 3.** Variations of transmission coefficient ($T$) and USL vs. the shell permittivity for the transmissive modes $T^{\pm1}$ and $T^0$ as shown in Fig. 1 for $N$=10.

According to these findings and based on the far-field expression derived in Eq. (5), a multi-objective genetic algorithm is applied to the analytical model to synthesize a scattering pattern with a maximum in the desired direction and minima in the others. This analytical procedure yields a reconfigurable metagrating capable of operating in three distinct states—State#A, State#B, and State#C—depending solely on the plasma frequencies of the two cores. These states correspond to beam steering toward $\varphi_t$ = +41°, $\varphi_t$ = 0°, and $\varphi_t$ = −41°, respectively, while maintaining the same geometrical configuration. For simplicity, the optimization in Eq. (5) was performed assuming $N$ = 10 unit-cells. While this number influences the beamwidth of each radiated lobe, it does not alter the beam direction, or compromise the optimization objectives. The optimal geometrical and physical parameters for this design are $r_1$=0.09$\lambda_0$, $r_2$=0.10$\lambda_0$, $d$=0.43$\lambda_0$, $P$=1.5$\lambda_0$, and $\varepsilon_r$=19.02, while the plasma frequencies in the three states are $f_{p1}/f_0$ = 4.5 and $f_{p2}/f_0$= 0.6 (State #A), $f_{p1}/f_0$ = $f_{p2}/f_0$ = 5.33 (State #B), and $f_{p1}/f_0$ = 0.6 and $f_{p2}/f_0$= 4.5 (State #C). For this structure, the scattered electric fields vs. the azimuthal angle $\varphi$ for the three states, with different main lobe directions, is depicted in Fig. 4, as derived by the analytical procedure with $N$=10.

A further optimization process based on full-wave simulations in CST is applied to the designed structure based on the analytical method. The optimization range has been chosen to be 50 % of the analytical values. The optimal parameters of the final design, which are close to those derived using the analytical method, are $r_1$=0.06$\lambda_0$, $r_2$=0.10$\lambda_0$, $d$=0.22$\lambda_0$, $P$=1.5$\lambda_0$, and $\varepsilon_r$=19.4, while the plasma frequencies in the three states are $f_{p1}/f_0$ = 5 and $f_{p2}/f_0$= 0.17 (State #A), $f_{p1}/f_0 = f_{p2}/f_0 = 6.5$ (State #B), and $f_{p1}/f_0 = 0.17$ and $f_{p2}/f_0 = 5$ (State #C). It is worth noting that the most significant variation occurs for the inter-cylinder distance $d$, mainly due to the approximate treatment of mutual coupling via Graf's addition theorem. Truncation of the infinite series and the neglect of higher-order mutual coupling can lead to inaccuracies, especially for closely spaced elements with a complex scattering response.

The reflection ($R$) and transmission coefficients ($T$) of the final design – calculated through full-wave simulations in unit-cell environment - are shown in Fig. 5(a). It should be noted that these coefficients correspond to different Floquet modes for each state and, therefore, to different propagation directions. Due to the symmetry of the structure, the reflection and transmission coefficients for states #A and #C are the same, and only one of them is reported. The results indicate that, for tilted radiation, approximately 92% of the normally-incident field is redirected to an angle of ±41°. For broadside radiation, this value is around 82%.

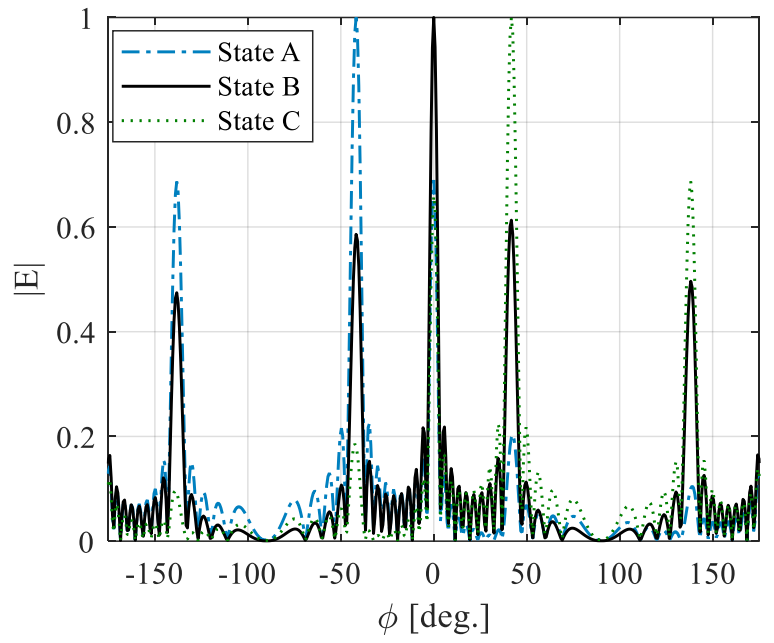

**Fig. 4.** Normalized scattered electric field on the horizontal plane of the proposed reconfigurable metagrating (analytical results with $N = 10$ and $\rho$=100 $\lambda_0$).

The normalized scattered electric field patterns and the electric field distributions of the designed metagrating for the three-states, obtained with full-wave simulations are plotted in Fig. 5(b) and Fig. 6, respectively. For comparison with Fig. 4, the results of Fig. 5(b) are also normalized to their maximum value. For the final design, the USL of tilted and broadside patterns are reduced to -13.15dB and -8.3dB, respectively. Additionally, Fig. 6 confirms high-efficiency beam steering in State#A and State#C, and almost a transparent response in State#B.

One of the most important parameters affecting the performance of plasma-based structures is plasma losses, which are directly related to the gas pressure [16]. To evaluate the robustness of the designed metagrating against plasma losses, the variations in the transmission coefficients and USL of all modes (0, ±1) as a function of gas pressure are depicted in Fig. 7. It can be observed that by changing the gas pressure ($0 \leq p \leq 3$ Torr), the transmission coefficients and USL for all states change by less than 20%.

Before concluding, we emphasize that the analysis developed in this work, based on the normalized plasma-to-operating frequency ratio, provides a general framework to evaluate metagrating performance independently of the absolute working frequency. In practice, the maximum achievable plasma frequency depends on experimental parameters such as gas type, pressure, and discharge configuration. Continuous discharges at low pressure typically allow operation in the tens of GHz range [21],[30]. Achieving higher plasma frequencies may require pulsed excitation to sustain the necessary electron densities, especially at low pressure. Given a specific plasma generation setup, the results presented here enable direct assessment of the maximum feasible operating frequency for practical implementation.

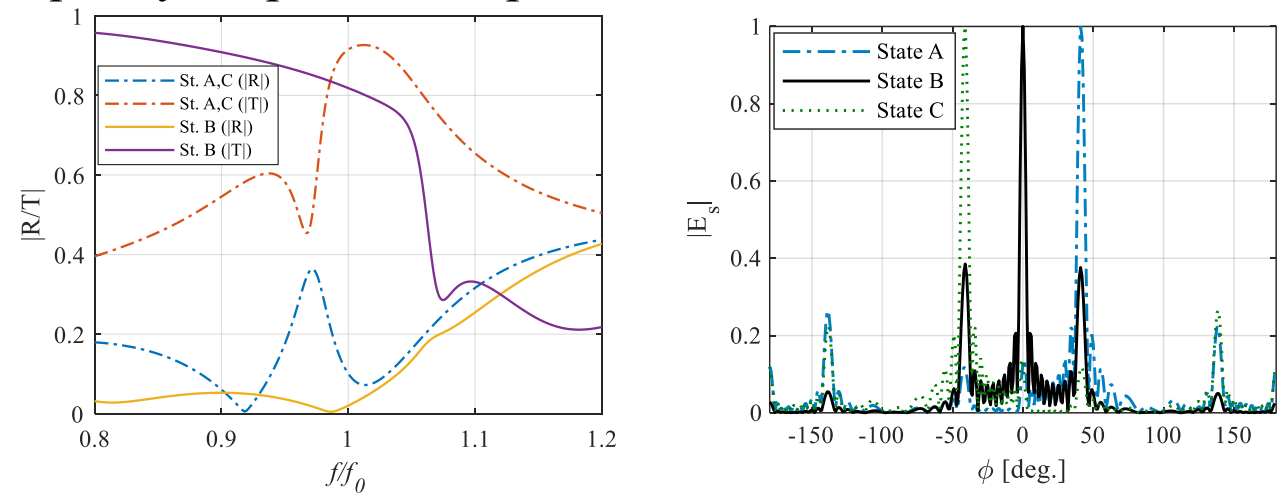

**Fig. 5.** (a) Transmission ($T$) and reflection ($R$) coefficients of the reconfigurable metagrating for its three operative states (full-wave results). The frequency axis is normalized to $f_0$. (b) Normalized scattered field patterns on the horizontal plane of the design single-layer tunable metagrating (full-wave results with $N$=10).

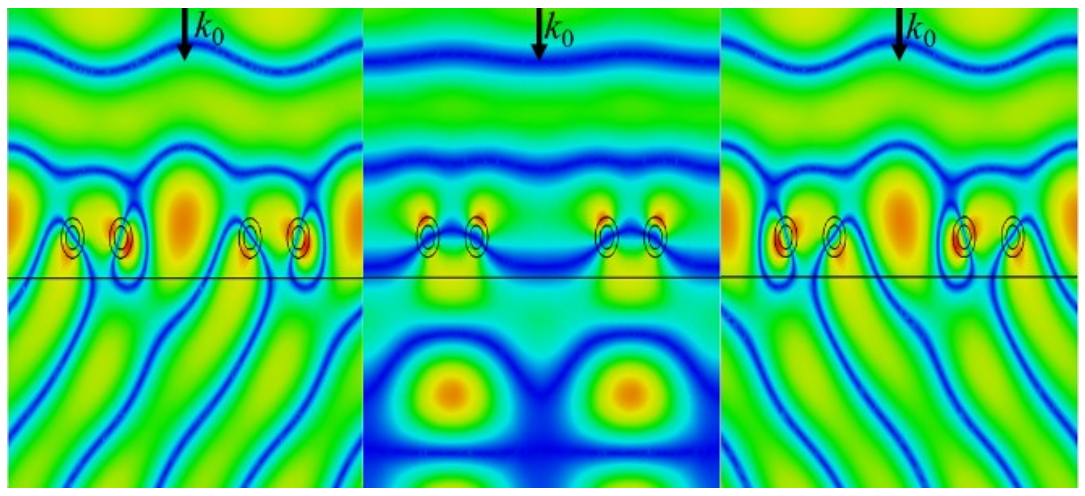

**Fig. 6.** Electric field distribution of the tunable metagrating in its three operating states.

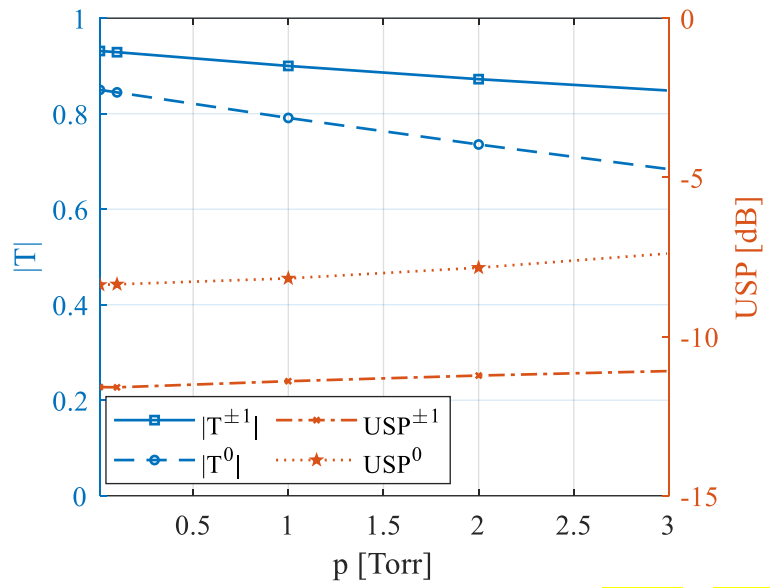

**Fig. 7.** Variations of transmission coefficients and USL (for $N$=10) vs. gas pressure for all modes.

## V. Conclusion

In this work, single-layer tunable metagratings was suggested by integrating plasma discharges with a high-refractive-index dielectric shell. It was confirmed that by tuning the plasma frequency through different discharge conditions, the main lobe direction could be switched between -41°, 0º, and +41°. The design displayed a high transmission coefficient, better than 82% for broadside radiation and 92% for a tilted pattern, confirming its effectiveness for adaptive beam steering applications. These findings open up potential applications in dynamic beamforming, reconfigurable antennas, and next-generation wireless communication systems, where real-time control of radiation patterns is essential.

## Acknowledgment

This work has been developed in the frame of the activities of the Project PULSE, funded by the European Innovation Council under the EIC Pathfinder Open 2022 program (protocol number 10109931). Project website is: https://www.pulse-pathfinder.eu/.